\documentclass[printer]{gOMS2e}

\doi{10.1080/1055.6788.2012.xxxxxx}
\issn{1029-4937}
\issnp{1055-6788}
\jvol{00} \jnum{00} \jyear{2012} \jmonth{November}

\usepackage{amssymb}
\usepackage{graphicx}
\usepackage{sidecap}

\usepackage{url}
\usepackage{graphicx}
\usepackage{times}
\usepackage{subfigure}
\usepackage{setspace}
\usepackage{latexsym}
\usepackage{algorithm}
\usepackage[noend]{algpseudocode}
\usepackage{wrapfig}
\usepackage{sidecap}
\usepackage{booktabs}

\def\clq{\mbox{\sc Clique}}

\def\clqh{\mbox{\sc CliqueHeu}}

\def\rowspace{-1pt}

\def\algospacing{1.0}

\newcommand{\algorithmroutine}[1]{-- \textit{#1}}

\begin{document}

\title{Fast Algorithms for the Maximum Clique Problem on \\ Massive Sparse Graphs}

\author{Bharath Pattabiraman$^{\rm \ast \ddag}$, Md. Mostofa Ali Patwary$^{\rm \ast \ddag}$, \\ Assefaw H. Gebremedhin$^{\rm \dag}$, Wei-keng Liao$^{\rm \ast}$, and Alok Choudhary$^{\rm \ast}$
\thanks{$^{\rm \ddag}$ Corresponding authors: \{bpa342,mpatwary\}@eecs.northwestern.edu \vspace{6pt}}
\\\vspace{6pt}  $^{\rm \ast}${\em{Northwestern University, Evanston, IL 60208}} \\$^{\rm \dag}${\em Purdue University, West Lafayette, IN 47907}
\\\vspace{6pt}
$^{\rm \ddag}$ Authors contributed equally \\ 
\received{November 2012} }

\maketitle

\begin{abstract}
The maximum clique problem is a well known NP-Hard problem with
applications in data mining, network analysis, informatics, and many other areas.
Although there exist several algorithms with acceptable runtimes for
certain classes of graphs, many of them are infeasible for massive graphs. 
We present a new exact algorithm that employs novel pruning techniques to 
very quickly find maximum cliques in large sparse graphs. 
Extensive experiments on several types of synthetic and 
real-world graphs show that our new algorithm is up to several orders of magnitude 
faster than existing algorithms for most instances. 
We also present a heuristic variant that runs orders of magnitude faster than 
the exact algorithm, while providing optimal or near-optimal solutions.
\end{abstract}

\section{Introduction}
\label{sec:intro}

A clique in an undirected graph is a subset of vertices in which every two vertices
are adjacent to each other. The {\em maximum} clique problem seeks to find 
a clique of the largest possible size in a given graph.

The maximum clique problem, and the related {\em maximal} clique and 
clique {\em enumeration} problems, find applications in diverse areas.
Some examples include  data mining \cite{Edachery99graphclustering,wang2009order,tsutomu2009clique}, 
information retrieval \cite{Augustson:1970:AGT:321607.321608}, 
social networks \cite{Fortunato_2010}, bioinformatics \cite{19566964},
computer vision \cite{Horaud:1989:SCT:68871.68875}, coding \cite{brouwer}, and 
economics \cite{RePEc:eee:csdana:v:48:y:2005:i:2:p:431-443}.
An example of its application can be given using data mining or information retrieval, where one needs to retrieve data that are considered similar based on some given metric. A graph is constructed with vertices corresponding to data items and edges connecting similar items. Finding a clique in such a graph gives a cluster of similar data. Such problems also arise in various other areas including identification and classification of new diseases based on symptom correlation \cite{Bonner:1964:CT:1662386.1662389}, pattern recognition \cite{1211348}, and bioinformatics \cite{19566964}.
More recently, the maximum clique problem has seen important applications in social network analysis, primarily in community detection \cite{Fortunato_2010,cite-key,5586496}. 
More examples of application areas for clique problems can be found in \cite{citeulike:4058448,Gutin2004}.

The maximum clique problem is NP-Hard \cite{Garey:1979:CIG:578533}.
Most exact algorithms for solving it employ some form of {\it branch-and-bound} approach. 
While branching systematically searches for all candidate solutions, bounding (also known as {\em pruning}) discards fruitless candidates based on a previously computed bound. An early example of a simple and  effective branch-and-bound algorithm for the maximum clique problem 
is one by Carraghan and Pardalos \cite{pardalos}.
More recently,  \"{O}sterg\.{a}rd \cite{ostergard} introduced 
an improved algorithm and demonstrated its relative advantages via computational experiments. 
Tomita and Seki \cite{citeulike:7905505}, and later, Konc and Janezic \cite{konc2007improved}
use upper bounds computed using vertex coloring to enhance the branch-and-bound approach. 
Other examples of branch-and-bound algorithms for the clique problem include 
\cite{Bomze99themaximum,SanSegundo,babel1990branch}.
Prosser \cite{prosser2012} has in a recent work compared various exact algorithms 
for the maximum clique problem.

An attractive feature of the algorithms of \cite{pardalos} and \cite{ostergard} is their 
simplicity in terms of ease of implementation.
However, their runtimes could  be infeasible  for very large graphs.
Furthermore, both algorithms as well as 
the algorithms from  \cite{citeulike:7905505} and \cite{konc2007improved}
are inherently sequential or otherwise difficult to parallelize. 
The ease with which an algorithm can be parallelized is important 
for handling large-scale graphs in emerging applications, where graphs with 
millions (or more) vertices are quite common \cite{kumar:extracting}.

In this paper, we present a new exact branch-and-bound
algorithm for the maximum clique problem that employs 
several new pruning strategies in addition to those in \cite{pardalos},
\cite{ostergard},  \cite{citeulike:7905505} and \cite{konc2007improved},
making it suitable for massive graphs.
We also present a heuristic that is based on similar pruning techniques as the exact algorithm
but runs much much faster---the heuristic follows just one of the ``paths" in the search space,
and as a result its complexity is nearly linear-time in the size of the graph, in contrast
to the exact algorithm whose worst-case complexity is exponential.
Both the exact algorithm and the heuristic are well-suited for parallelization.
The algorithms are discussed in detail in Section~\ref{sec:algorithms}.

In Section~\ref{sec:experiments} we present an extensive experimental analysis
comparing the performance of our algorithms with the algorithm of
Carraghan and Pardalos \cite{pardalos}, the algorithm of \"{O}sterg\.{a}rd \cite{ostergard}
and the algorithm of Konc and Janezic \cite{konc2007improved}.
The workings of the latter three algorithms is reviewed in Section~\ref{sec:relatedwork}.
Our testbed includes large-scale real-world graphs drawn from various application domains,
large-scale synthetic graphs representing various structures,
and DIMACS benchmark graphs.
The new exact algorithm is found to be up to orders of magnitude faster on large,
sparse graphs and of comparable runtime on denser graphs.
The heuristic in turn is found to run several orders of magnitude faster than the exact algorithm,
while delivering solutions that are optimal or near-optimal for most cases.
We have made our implementations publicly available at 
\url{http://cucis.ece.northwestern.edu/projects/MAXCLIQUE/}.

\section{Related Previous Algorithms}
\label{sec:relatedwork}

Given a simple undirected graph $G$, the maximum clique can clearly be obtained by enumerating 
{\em all} of the cliques present in it and picking the largest of them.
Carraghan and Pardalos \cite{pardalos} introduced a simple-to-implement
algorithm that avoids enumerating all cliques and instead
works with a significantly reduced partial enumeration.
The reduction in enumeration is achieved via 
a {\em pruning} strategy which reduces the search space tremendously.
The algorithm works by
performing at each step $i$, a {\em depth first search} from vertex $v_i$, 
where the goal is to find the largest clique containing the vertex $v_i$.
At each {\em depth} of the search, the algorithm compares the number of remaining
vertices that could potentially constitute a clique containing vertex $v_i$
against the size of the largest clique encountered thus far.
If that number is found to be smaller, the algorithm backtracks (search is pruned).

\"{O}sterg\.{a}rd \cite{ostergard} devised an algorithm that incorporated an additional 
pruning strategy to the one by Carraghan and Pardalos.
The opportunity for the new pruning strategy is created by {\em reversing} the order in which the search is done by
the Carraghan-Pardalos algorithm. This allows for an additional pruning with the help of
some auxiliary bookkeeping. 
Experimental results in \cite{ostergard} showed that the \"{O}sterg\.{a}rd
algorithm is faster than the one by Carraghan-Pardalos on random and DIMACS benchmark graphs \cite{dimacs}.
However, the new pruning strategy used in this algorithm is intimately tied to the order in which vertices are processed, introducing an inherent sequentiality into the algorithm.

A number of existing branch-and-bound algorithms for maximum clique
use a vertex-coloring of the graph to obtain an upper bound on the maximum clique. 
A {\em vertex-coloring} of a graph is an assignment of colors to vertices such that 
a pair of adjacent vertices receive different colors. Clearly, the number of colors used gives 
an upper bound on the maximum clique of the graph, which can be used to reduce the search space. 
A popular and recent algorithm based on this idea is 
the algorithm of Tomita and Seiku \cite{citeulike:7905505} (known as MCQ). 
More recently, Konc and Janezic \cite{konc2007improved} presented 
an improved version of MCQ, known as MaxCliqueDyn (MCQD and MCQD+CS), that
involves the use of tighter, computationally more expensive upper bounds 
applied on a fraction of the search space.


\section{The New Algorithms}
\label{sec:algorithms}

We describe in this section new algorithms that overcome the aforementioned shortcomings---the new algorithms 
use additional pruning strategies, maintain simplicity, and avoid sequential computational order.
Before going into the details of the algorithms, 
we introduce a few notations used throughout the paper. 
We identify the $n$ vertices of the input graph $G=(V,E)$ as $\{v_1, v_2, \ldots, v_n\}$.  
The set of vertices adjacent to a vertex $v_i$, the set of its neighbors, is denoted by $N(v_i)$.
And the cardinality of $N(v_i)$, its degree, is denoted by $d(v_i)$.  

\subsection{The Exact Algorithm}
\label{subsec:exact}

The maximum clique in a graph can be found by computing the largest clique containing each vertex and picking the largest among these. 
A key element of our exact algorithm is that during the search for the largest clique containing a given vertex, vertices that cannot form cliques larger than the current maximum 
clique are {\em pruned}, in a hierarchical fashion. 
The method is outlined in detail in Algorithm \ref{alg:mClq}. 
Throughout the algorithm, the variable $max$ stores the size of the maximum clique found 
thus far. Initially it is set to be equal to the lower bound $lb$ provided as an input parameter,
and it gives the maximum clique size when the algorithm terminates.

To obtain the largest clique containing a vertex $v_i$, it is sufficient to consider only the neighbors of $v_i$. 
The main routine {\sc MaxClique} thus generates  for each vertex $v_i \in V$ a set $U \subseteq N(v_i)$ 
(neighbors of $v_i$ that survive pruning) and calls the subroutine \clq\ on $U$.  
The subroutine \clq\ goes through every relevant clique containing $v_i$ 
in a recursive fashion and returns the largest. 
The subroutine is similar to the Carraghan-Pardalos algorithm \cite{pardalos}.
We use $size$ to maintain the size of the clique found at any point through the recursion.
Since we start with a clique of just one vertex, the value of $size$ is set to be one initially 
when the subroutine \clq\ is called (Line \ref{subCall} of Algorithm \ref{alg:mClq}).

\begin{figure}[t]
\begin{center}

\begin{minipage}{0.75\textwidth}
\begin{algorithm}[H]
\begin{spacing}{\algospacing}
{
\small
\caption{{\protect\small Algorithm for finding the maximum clique of a given graph.
{\it Input}: Graph $G = \left (V, E\right )$, lower bound on clique $lb$ (default, 0). 
{\it Output}: Size of maximum clique.}}

\label{alg:mClq}
\begin{algorithmic}[1]
\Procedure {MaxClique}{$G=\left (V,E\right )$, $lb$}
\State $max \leftarrow lb$
\For{$i:1$ to $n$}
\If{$d(v_i) \ge max$} \Comment{{\footnotesize Pruning 1}} \label{pr1}
\State $U \leftarrow \emptyset$
\For{each $v_j \in N(v_i)$}
\If{$j > i$} \Comment{{\footnotesize Pruning 2}} \label{prOld}
\If{$d(v_j) \ge max$} \Comment{{\footnotesize Pruning 3}} \label{pr2}
\State $U \leftarrow U \cup \{v_j\}$ 
\EndIf
\EndIf
\EndFor

\State \textsc{Clique}$(G, U, 1)$ \label{subCall}
\EndIf

\EndFor
\EndProcedure
\end{algorithmic}
}
\rule{1\textwidth}{.1mm}\\
\algorithmroutine{Subroutine}
{
\begin{algorithmic}[1]
\Procedure {Clique}{$G=\left (V,E\right )$, $U$, $size$}

\If{$U = \emptyset$}
\If{$size > max$}
\State $max \leftarrow size$
\EndIf
\State {\bf return}
\EndIf

\While{$\left|U\right| > 0$}
\If{$size + \left|{U}\right| \le max$} \Comment{{\footnotesize Pruning 4}} \label{pr3}
\State {\bf return} 
\EndIf

\State Select any vertex $u$ from $U$ 

\State $U \leftarrow U \setminus \{u\} $
\State $N'(u):= \{w | w \in N(u) \wedge d(w) \ge max\}$  \Comment{{\footnotesize Pruning 5}} \label{pr4}
\State \textsc{Clique}$( G, U \cap N'(u), size + 1)$
\EndWhile

\EndProcedure
\end{algorithmic}
}
\end{spacing}
\end{algorithm}
    \end{minipage}
\end{center}
\end{figure}

Our algorithm consists of several pruning steps.
The pruning in Line \ref{pr1} of {\sc Max-} 
{\sc Clique} (Pruning 1)
filters vertices having strictly fewer neighbors than the size of the maximum clique already computed. These vertices can be safely ignored, since even if a clique were to be found, its size would not be larger than $max$.
While forming the neighbor list $U$ for a vertex $v_i$, we include only those of $v_i$'s 
neighbors for which the largest clique containing them has not been found (Line \ref{prOld}, Pruning 2),
 to avoid recomputing previously found cliques.  
Furthermore, the pruning in Line \ref{pr2} (Pruning 3)
excludes vertices $v_j \in N(v_i)$  that have degree less than the current value of $max$, since any such vertex could not form a clique of size larger than $max$.
The pruning strategy in Line \ref{pr3} of subroutine \clq\ (Pruning 4)
checks for the case where even if all vertices of $U$ were added to get a clique, its size would not exceed that of the largest clique encountered so far in the search, $max$. 
The pruning in Line 11 of \clq\ (Pruning 5)
reduces the number of comparisons needed to generate the intersection set in Line 12.
Pruning 4 is used in most existing algorithms, whereas pruning steps 1, 2, 3 and 5 are new.

\subsection{The Heuristic}
\label{subsec:heuristic}

The exact algorithm examines for every vertex $v_i$ all relevant cliques containing the vertex $v_i$
in order to determine the clique of maximum size among them.
Our heuristic speeds up this process by instead examining only a subset of the relevant cliques. 
  
The heuristic is presented in Algorithm \ref{alg:mClqHeu}. The main routine is very similar to the main routine in Algorithm \ref{alg:mClq}. The subroutine \clqh\ considers only the {\em maximum degree} neighbor at each step instead of recursively considering all neighbors from the set $U$. Since we are looking for the largest clique containing each vertex, the maximum degree vertex is more likely to be a member of the largest clique compared to the other vertices. The effect of choosing the maximum degree vertex as opposed to any random vertex will be analyzed in Section~\ref{sec:exp-heuristic}.
We note that Turner \cite{Turner88} uses an algorithm similar in spirit to the subroutine of Algorithm \ref{alg:mClqHeu} in his coloring algorithm. 

\begin{figure}[t]
\begin{center}

\begin{minipage}{0.68\textwidth}
\begin{algorithm}[H]
\begin{spacing}{\algospacing}
{
\small
\caption{{\protect\small Heuristic for finding the maximum clique in a graph.
{\it Input}: Graph $G = \left (V, E\right )$. {\it Output}: Approximate size of maximum clique.}}
\label{alg:mClqHeu}

\begin{algorithmic}[1]
\Procedure {MaxCliqueHeu}{$G=\left (V,E\right )$}
\For{$i:1$ to $n$}
\If{$d(v_i) \ge max$} \Comment{{\footnotesize Pruning 1}}
\State $U \leftarrow \emptyset$
\For{each $v_j \in N(v_i)$}
\If{$d(v_j) \ge max$} \Comment{{\footnotesize Pruning 3}}
\State $U \leftarrow U \cup \{v_j\}$ 
\EndIf
\EndFor

\State \textsc{CliqueHeu}$(G, U, 1)$
\EndIf

\EndFor
\EndProcedure
\end{algorithmic}
}
\rule{1\textwidth}{.1mm}\\
\algorithmroutine{Subroutine}

{
\label{alg:clqHeu}
\begin{algorithmic}[1]
\Procedure {CliqueHeu}{$G=\left (V,E\right )$, $U$, $size$}

\If{$U = \emptyset$}
\If{$size > max$}
\State $max \leftarrow size$
\EndIf
\State {\bf return}
\EndIf


\State Select a vertex $u \in U$ of maximum degree in $G$ \label{maxDsel}
\State $U \leftarrow U \setminus \{u\} $
\State $N'(u):= \{w | w \in N(u) \wedge d(w) \ge max\}$  \Comment{{\footnotesize Pruning 5}} \label{pr4}
\State \textsc{CliqueHeu}$( G, U \cap N'(u), size + 1)$


\EndProcedure
\end{algorithmic}
}
\end{spacing}
\end{algorithm}
    \end{minipage}
\end{center}
\end{figure}

\subsection{Complexity}
\label{subsec:complexity}

The exact algorithm, Algorithm \ref{alg:mClq}, examines for every vertex $v_i$ all candidate cliques containing the vertex $v_i$ in its search for the largest clique. Its time complexity is exponential in the worst case. The heuristic, Algorithm \ref{alg:mClqHeu}, loops over the $n$ vertices, each time possibly
calling the subroutine \clqh, which effectively is a loop that runs until the set $U$ is empty. 
Clearly, $|U|$ is bounded by the max degree $\Delta$ in the graph.  
The subroutine also includes the computation of a neighbor list, whose runtime is bounded by 
$O(\Delta)$.
Thus, the time complexity of the heuristic is bounded by $O(n\cdot \Delta^{2})$.

\section{Experiments and Results}
\label{sec:experiments}

We present in this section experimental results comparing the performance of our algorithm
with the algorithms of Carraghan-Pardalos \cite{pardalos}, 
\"{O}sterg\.{a}rd algorithm \cite{ostergard}, and
Konc and Janezik \cite{konc2007improved}.

We implemented the algorithm of \cite{pardalos} ourselves, whereas for the algorithm of \cite{ostergard},  we used the publicly available {\it cliquer} source code \cite{cliquer},
and similarly, for the algorithm of \cite{konc2007improved}
we used the code {\it MaxCliqueDyn} (MCQD, available at \url{http://www.sicmm.org/~konc/maxclique/}). Among the variants available in MCQD, we report results on the best-performing variant, the variant called
MCQD+CS (that uses improved coloring and dynamic sorting). 

All our experiments are performed on a Linux workstation running 64-bit version Red Hat Enterprise Linux Server release 6.2, with a 2.00 GHz Intel Xeon E7540 processor. Our implementations are all in C++, and the codes are compiled using gcc version 4.4.6 with -O3 optimization. 

\subsection{Test Graphs}

Our testbed is grouped in three categories.

\subsubsection{Real-world graphs} 
Under this category, we consider 10 graphs (downloaded from the 
University of Florida Sparse Matrix Collection  \cite{Davis97theuniversity}) that originate
from various real-world applications. Table~\ref{tab:real-graphs} gives a quick overview of the graphs and their origins.
  
\begin{table}
\centering
\caption{Overview of real-world graphs in the testbed and their origins.}
\label{tab:real-graphs}
\begin{tabular}{ll}
{\bf Graph} & {\bf Description} \\ \hline \hline
{\it cond-mat-2003} \cite{Newman06042004} & A collaboration network of scientists posting preprints 
on \\ & the condensed matter archive at www.arxiv.org in the period \\ & between January 1, 1995 and June 30, 2003. \\ \hline
{\it email-Enron} \cite{Leskovec:2005:GOT:1081870.1081893} & A communication network representing
email exchanges. \\
& Nodes are email addresses and there is a directed edge from \\ & node $i$ to node $j$ if at least one email is sent from $i$ to $j$. \\ \hline
{\it dictionary28} \cite{pajek2006} & Pajek network of words. \\ \hline
{\it Fault\_639} \cite{Ferronato20083922} & A structural problem discretizing a faulted gas reservoir with \\
& tetrahedral Finite Elements and triangular Interface Elements. \\ \hline
{\it audikw\_1} \cite{Davis97theuniversity} & An automotive crankshaft model of TETRA elements. \\ \hline
{\it bone010} \cite{vanRietbergen199569} &
A detailed micro-finite element (micro-FE) model of bones \\
& representing the porous bone micro-architecture. \\ \hline
{\it af\_shell} \cite{Davis97theuniversity}  & A sheet metal forming simulation network. \\ \hline
{\it as-Skitter} \cite{Leskovec:2005:GOT:1081870.1081893} & An Internet topology graph from trace routes run daily in 2005. \\ \hline 
{\it roadNet-CA} \cite{Leskovec:2005:GOT:1081870.1081893} & A road network of California.
Nodes represent intersections \\ & and endpoints and edges represent the roads connecting the \\ & intersections or endpoints. \\    \hline
{\it kkt\_power} \cite{Davis97theuniversity} & An Optimal Power Flow (nonlinear optimization) network. \\\hline
\end{tabular}
\end{table}

\subsubsection{Synthetic Graphs} 
In this category we consider 15 graphs generated using 
the R-MAT algorithm \cite{Chakrabarti:2006:GML:1132952.1132954}. The graphs
are subdivided in three categories depending on the structures they represent.
\begin{itemize}
\item {\bf Random graphs} (5 graphs) -- Erd\H{o}s-Renyi random  graphs generated using R-MAT 
with the parameters (0.25, 0.25, 0.25, 0.25).  The graphs are denoted with prefix {\it rmat\_er}.
\item {\bf Skewed Degree, Type 1 graphs} (5 graphs) -- graphs generated using R-MAT with the parameters 
(0.45, 0.15, 0.15, 0.25). Denoted with prefix {\it rmat\_sd1}.
\item {\bf Skewed Degree, Type 2 graphs} (5 graphs) --  graphs generated using R-MAT with the parameters (0.55, 0.15, 0.15, 0.15). Denoted with prefix {\it rmat\_sd2}.
\end{itemize}

\subsubsection{DIMACS graphs} 
This last category consists of 5 graphs selected from the Second DIMACS Implementation Challenge \cite{dimacs}. 

The DIMACS graphs  are an established benchmark for the maximum
clique problem, but they are of rather limited size and variation. 
In contrast, the real-work networks included  in category 1 of the testset
and the synthetic (RMAT) graphs in category 2
represent a wide spectrum of large graphs posing varying degrees of difficulty for testing the algorithms. 
The {\it rmat\_er} graphs have {\it normal} degree distribution, whereas the {\it rmat\_sd1} and {\it rmat\_sd2} graphs have skewed degree distributions and contain many dense local subgraphs.
 The {\it rmat\_sd1} and {\it rmat\_sd2} graphs differ primarily in the magnitude of maximum vertex degree they contain; the {\it rmat\_sd2} graphs have much higher maximum degree. 
Table \ref{tab:struc-graphs} lists basic structural information (the number of vertices, 
number of edges and the maximum degree) about all 30 of the test graphs.

\begin{table}[t]
\small
\centering
\caption{Structural properties (the number of vertices, $|V|$; edges, $|E|$; and the maximum degree, $\Delta$) of the graphs, $G$ in the testbed:  
DIMACS Challenge graphs (upper left); UF Collection (lower and middle left);
RMAT graphs (right).}  
\label{tab:struc-graphs}
\begin{tabular}{l@{\hspace{5pt}}r@{\hspace{5pt}}r@{\hspace{5pt}}r@{\hspace{5pt}}|@{\hspace{5pt}}l@{\hspace{5pt}}r@{\hspace{5pt}}r@{\hspace{5pt}}r}

\toprule\toprule

$G$ & $|V|$ & $|E|$ & $\Delta$ & $G$ & $|V|$ & $|E|$ & $\Delta$ \\ \hline \hline
{\it cond-mat-2003} & 31,163	& 120,029	 & 202 &	{\it rmat\_sd1\_1} &    131,072 &    1,046,384 & 407           \\ \vspace*{\rowspace}
{\it email-Enron} & 36,692	 & 183,831 &	1,383  &	{\it rmat\_sd1\_2} &    262,144 &    2,093,552 &   558    \\ \vspace*{\rowspace}
{\it dictionary28} & 	52,652 &	89,038 &	38  &		{\it rmat\_sd1\_3} &    524,288 &    4,190,376 &    618  \\ \vspace*{\rowspace}
{\it Fault\_639} &    638,802 &    13,987,881 &    317 &  {\it rmat\_sd1\_4} &    1,048,576 &    8,382,821 &  802    \\ \vspace*{\rowspace}
{\it audikw\_1} &    943,695 &    38,354,076 &    344  &	 {\it rmat\_sd1\_5} &    2,097,152 &    16,767,728 &    1,069   \\

\midrule
\vspace*{\rowspace}
 {\it bone010} &    986,703 &    35,339,811 &    80 & 
{\it rmat\_sd2\_1} &    131,072 &    1,032,634 &    2,980        \\ \vspace*{\rowspace}
{\it af\_shell10} &    1,508,065 &    25,582,130 &    34 &  
{\it rmat\_sd2\_2} &    262,144 &    2,067,860 &    4,493      \\ \vspace*{\rowspace}
{\it as-Skitter} &    1,696,415 &    11,095,298 &  35,455 &
{\it rmat\_sd2\_3} &    524,288 &    4,153,043 &    6,342     \\ \vspace*{\rowspace}
{\it roadNet-CA} &    1,971,281 &    2,766,607 &    12 & 
{\it rmat\_sd2\_4} &    1,048,576 &    8,318,004 &    9,453      \\ \vspace*{\rowspace}	
{\it kkt\_power} &    2,063,494 &    6,482,320 &    95  &   
{\it rmat\_sd2\_5} &    2,097,152 &    16,645,183 &    14,066    \\

\midrule
\vspace*{\rowspace}
{\it rmat\_er\_1} &    131,072 &    1,048,515 &    82  &   
{\it hamming6-4} & 64 &    704 &    22 	 \\ \vspace*{\rowspace}
{\it rmat\_er\_2} &    262,144 &    2,097,104 &    98 &  
{\it johnson8-4-4} & 70 &    1,855 &    53   \\ \vspace*{\rowspace} 
{\it rmat\_er\_3} &    524,288 &    4,194,254 &    94  &  
{\it keller4} &    171 &    9,435 &    124  \\ \vspace*{\rowspace}   
{\it rmat\_er\_4} &    1,048,576 &    8,388,540 &    97  & 
{\it c-fat200-5} &    200 &    8,473 &    86  \\ \vspace*{\rowspace} 
{\it rmat\_er\_5} &    2,097,152 &    16,777,139 &    102 & 
{\it brock200\_2} &    200 &    9,876 &    114 \\

\bottomrule\bottomrule
\end{tabular}
\end{table}

\subsection{Results}

\label{sec:exp-results}

\begin{table}[!hbt]

\small
\centering
\caption{Comparison of runtimes (in seconds) of algorithms \cite{pardalos} ({\it CP}),
\cite{ostergard} ({\it cliquer}) and \cite{konc2007improved} ({\it MCQD+CS})
with the time taken by our new exact algorithm ($\tau_{new-exact}$) for the graphs in the testbed, with the fastest (marked in bold) for each case. An asterisk (*) indicates that the algorithm did not terminate within 25,000 seconds for that instance. A hyphen (-) indicates the publicly available implementation by the authors of algorithm terminated due to the graph being too large for the implementation to handle. 
$\omega$ denotes the maximum clique size, $\omega_{new-heuristic}$, the maximum clique size returned by our heuristic and $\tau_{new-heuristic}$, its runtime. For the graph {\it rmat\_sd2\_5}, none of the algorithms computed the maximum clique size in a reasonable time;
the entry is marked $N$, denoting ``Not Known'').}
\label{tab:timings}
\begin{tabular}{l@{\hspace{6pt}}r@{\hspace{6pt}}|@{\hspace{6pt}}r@{\hspace{6pt}}r@{\hspace{6pt}}r@{\hspace{6pt}}r@{\hspace{6pt}}|@{\hspace{6pt}}r@{\hspace{6pt}}r}

\toprule\toprule
           &  & & &   & $\tau_{new-}$ & $\omega_{new-}$& $\tau_{new-}$\\
Graph           & $\omega$ & $\tau_{CP}$    & $\tau_{cliquer}$  & $\tau_{MCQD+CS}$  & $_{exact}$ & $_{heuristic}$& $_{heuristic}$\\
\hline \hline
{\it cond-mat-2003} 	& 	25 	& 	4.875 		&  	11.17		&	2.41		&	{\bf 0.011}		&	25 		& 	$<$0.01 	\\
{\it email-Enron} 	& 	20 	& 	7.005		& 	15.08 		&	3.70		& 	{\bf 0.998}		&	18 		& 	0.261	\\ 
{\it dictionary28} 	& 	26 	& 	7.700 		&	32.74 		&	7.69		&	{\bf $<$0.01}	&	26 		&	$<$0.01	\\	
{\it Fault\_639}		&	18	&	14571.20		&	4437.14		&	-		&	{\bf 20.03}		&	18		&	5.80 		\\
{\it audikw\_1}		&	36	&	*			&	9282.49		&	-		&	{\bf 190.17}	&	36		&	58.38 	\\
{\it bone010}		&	24	&	*			&	10002.67		&	-		&	{\bf 393.11}	&	24		&	24.39 	\\ 
{\it af\_shell10}		&	15	&	*			&	21669.96		&	-		&	{\bf 50.99}		&	15		&	10.67 	\\
{\it as-Skitter}		&	67	&	24385.73		&	*			&	-		&	{\bf 3838.36}	&	66		&	27.08 	\\ 
{\it roadNet-CA}	&	4	&	*			&	*			&	-		&	{\bf 0.44}		&	4		&	0.08 		\\ 
{\it kkt\_power}		&	11	&	*			&	*			&	-		&	{\bf 2.26}		&	11		&	1.83 		\\ 
\midrule
{\it rmat\_er\_1}		&	3	&	256.37		&	215.18		&	49.79	&	{\bf 0.38}		&	3		&	0.12 		\\
{\it rmat\_er\_2}		&	3	&	1016.70		&	865.18		&	-		&	{\bf 0.78}		&	3		&	0.24 		\\
{\it rmat\_er\_3}		&	3	&	4117.35		&	3456.39		&	-		&	{\bf 1.87}		&	3		&	0.49 		\\
{\it rmat\_er\_4}		&	3	&	16419.80		&	13894.52		&	-		&	{\bf 4.16}		&	3		&	1.44 		\\
{\it rmat\_er\_5}		&	3	&	*			&	*			&	-		&	{\bf 9.87}		&	3		&	2.57 		\\
\midrule
{\it rmat\_sd1\_1}	&	6	&	225.93		&	214.99		&	50.08	&	{\bf 1.39}		&	6		&	0.45 		\\
{\it rmat\_sd1\_2}	&	6	&	912.44		&	858.80		&	-		&	{\bf 3.79}		&	6		&	0.98 		\\ 
{\it rmat\_sd1\_3}	&	6	&	3676.14		&	3446.02		&	-		&	{\bf 8.17}		&	6		&	1.78 		\\ 
{\it rmat\_sd1\_4}	&	6	&	14650.40		&	13923.93		&	-		&	{\bf 25.61}		&	6		&	4.05 		\\ 
{\it rmat\_sd1\_5}	&	6	&	*			&	*			&	-		&	{\bf 46.89}		&	6		&	9.39 		\\ 
\midrule
{\it rmat\_sd2\_1}	&	26	&	427.41		&	213.23		&	{\bf 48.17}	&	242.20		&	26		&	32.83 	\\ 
{\it rmat\_sd2\_2}	&	35	&	4663.62		&	{\bf 851.84}	&	-		&	3936.55		&	35		&	95.89 	\\ 
{\it rmat\_sd2\_3}	&	39	&	13626.23		&	{\bf 3411.14}	&	-		&	10647.84		&	37		&	245.51 	\\ 
{\it rmat\_sd2\_4}	&	43	&	*			&	{\bf 13709.52}	&	-		&	*			&	42		&	700.05 	\\
{\it rmat\_sd2\_5}	&	N	&	*			&	*			&	-		&	*			&	51		&    1983.21 	\\
\midrule
{\it hamming6-4}	&	4	&	{\bf $<$0.01}	&	{\bf $<$0.01}	&{\bf $<$0.01}	&	{\bf $<$0.01}	&	4		&	$<$0.01 	\\
{\it johnson8-4-4}	&	14	&	0.19			&	{\bf $<$0.01}	&{\bf $<$0.01}	&	0.23			&	14		&	$<$0.01 	\\
{\it keller4}		&	11	&	22.19		&	0.15			&	{\bf 0.02}	&	23.35		&	11		&	$<$0.01 	\\
{\it c-fat200-5}		&	58	&	0.60			&	0.33			&	{\bf 0.01}	&	0.93			&	58		&	0.04 		\\
{\it brock200\_2}	&	12	&	0.98			&	0.02			&{\bf $<$0.01}	&	1.10			&	10		&	$<$0.01 	\\
\bottomrule\bottomrule
\end{tabular}

\end{table}

Table \ref{tab:timings} shows the size of the maximum clique ($\omega$) and the runtimes  of our exact algorithm and the algorithms of Caraghan and Pardalos \cite{pardalos} (CP), \"{O}sterg\.{a}rd \cite{ostergard} ({\it cliquer}) and Konc and Janezic \cite{konc2007improved} 
(MCQD+CS) for all the graphs in the testbed. 
The last two columns show the results of our heuristic---the size of the maximum clique 
returned  and its runtime. 

In Section~\ref{sec:exp-exact} and Section~\ref{sec:exp-heuristic},
we discuss our observations from this table
for the exact algorithm and the heuristic, respectively, but before that we 
briefly comment on our experience in using the {\it MaxCliqueDyn} code.
Unfortunately, the code failed to execute most of the  
large instances in our testbed, including the majority of the RMAT and real-world instances,
due to memory management issues in the code. 
The entries in Table~\ref{tab:timings} marked with hyphen (-) show instances
for which the code was aborted due to excessive memory usage.
Even for the instances it eventually run successfully, we had to first make modifications to the graph reader to make it able to handle graphs with multiple connected components.


\subsubsection{Exact algorithms}
\label{sec:exp-exact}

As expected, our exact algorithm gave the same size of maximum clique as the other
three algorithms for all test cases. 
In terms of runtime,  its relative performance compared to the other three varied
in accordance with the advantages afforded by the various pruning steps.

\begin{figure}
  \centering
    \includegraphics[scale=0.45]{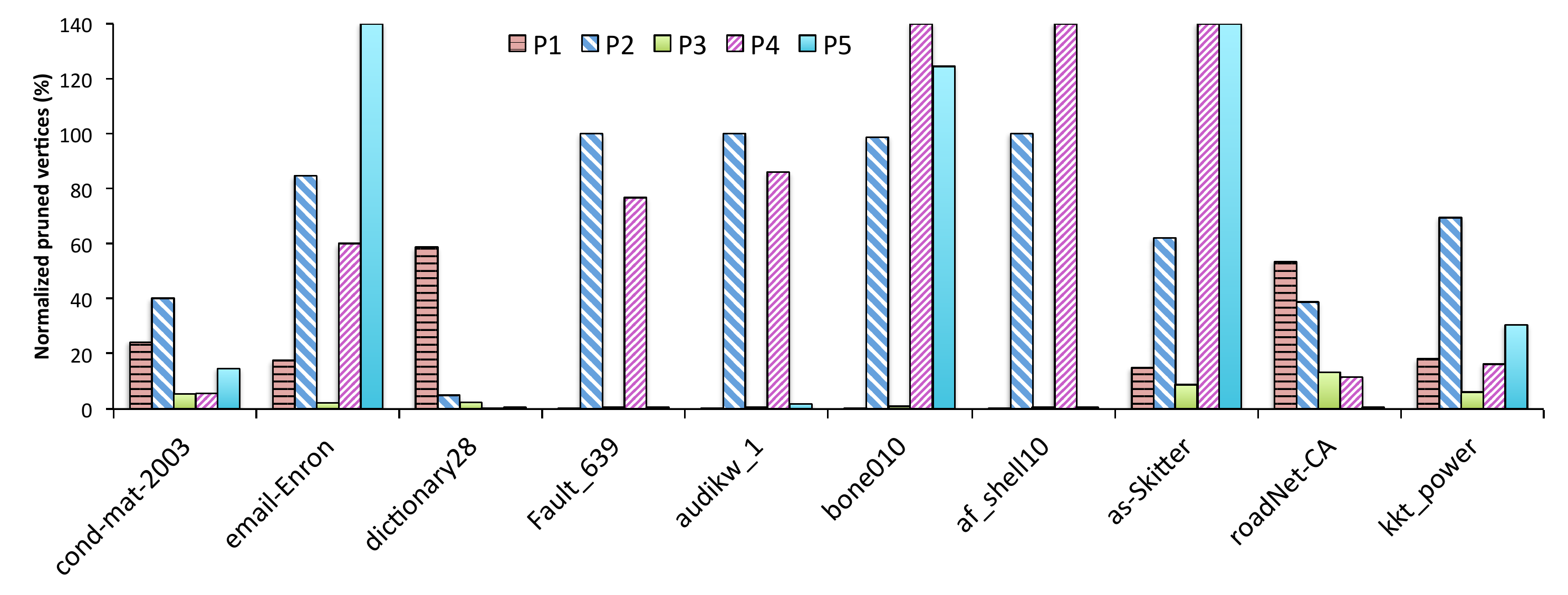}
\caption{Number of ``pruned" vertices in the various pruning steps normalized
by the number of edges in the graph (in percents) for the test graphs in category 1 (we cut few bars reachining 140\% as their correspnding values are much higher).}
\label{fig-pruningplot}
\end{figure}

{\bf Analysis of pruning steps.  }
Vertices that are discarded by Pruning 1 are skipped in the main loop of the algorithm, and the largest cliques containing them are not computed. Pruning 2 avoids re-comuting previously computed cliques in the neighborhood of a vertex. In the absence of Pruning 1, the number of vertices pruned by Pruning 2 would be bounded by the number of edges in the graph (note that this is more than the total number of vertices in the graph). While Pruning 3 reduces the size of the input set on which the maximum clique is to be computed, Pruning 5 brings down the time taken to generate the intersection set in Line 12 of the subroutine. 
Pruning 4 corresponds to back tracking. Unlike Pruning steps 1, 2, 3 and 5, Pruning 4
is used  by all three of the other algorithms in our comparison.

One of the strengths of our algorithm is its ability to take advantage of pruning in multiple
steps in a hierarchical fashion, allowing for opportunities for one or more of the steps to
kick in and impact performance.
In Figure~\ref{fig-pruningplot} we show the number of vertices discarded by all
the  pruning steps of the exact algorithm normalized by the total number of edges
in a graph for the real-world graphs (category 1) in the testbed. We cut few bars reachining
140\% as their correspnding values are much higher.
It can be seen for these graphs pruning steps 2 and 5 in particular discard a large 
percentage of vertices, potentially resulting in large runtime savings.
The general behavior of the pruning steps Pruning 1, 2, 3 and 5 for the synthetic graphs 
{\em rmat\_er} and {\em rmat\_sd1} was observed to be somewhat similar to that depicted in 
Figure~\ref{fig-pruningplot} for the real-world graphs. 
In contrast, for the DIAMCS graphs, the number of vertices pruned in steps
Pruning 1, 3 and 5 were observed to be zero; the numbers in the step 
Pruning 2 were nonzero, but relatively modest.
In the Appendix, we provide a complete tabulation of the raw numbers for the pruned vertices
in all the steps for all the graphs in the testbed.

As a result  of the differences seen in the effects of the pruning steps, as discussed below,
the runtime performance of our algorithm (seen in Table \ref{tab:timings}) compared
to the other three algorithms varied in accordance with the difference in the structures represented 
by the different categories of graphs in the testbed.

{\bf Real-world Graphs. }
For most of the graphs in this category, it can be seen that our algorithm runs several orders of magnitude faster than the other three, mainly due to the large amount of pruning the algorithm enforced. 
For the graphs {\em Fault\_639}, {\em audikw\_1} and {\em af\_shell10}, 
Prunings 1, 3 and 5 had relatively small impact, 
whereas, Pruning 2 makes a huge impact. 
The number of vertices pruned in steps Pruning 1 and 3 varied among the graph {\em within} the
category, ranging from 0.001\% for {\it af\_shell} to a staggering 97\% for {\it as-Skitter} 
for the step Pruning 1 (see the table in the Appendix for details). 

{\bf Synthetic Graphs. }
For the synthetic graph types {\it rmat\_er} and {\it rmat\_sd1}, our algorithm clearly outperforms 
the other three by a few orders of magnitude in all cases. 
This is also primary due to the high number of vertices discarded by the new pruning steps. 
In particular, for {\it rmat\_sd1} graphs, between 30 to 37\% of the vertices are pruned just in the step Pruning 1. 
For the {\it rmat\_sd2} graphs, which have relatively larger maximum clique and higher maximum degree than the {\it rmat\_sd1} graphs, our algorithm is observed to be faster than 
CP but slower than {\em cliquer}. 

{\bf DIMACS Graphs. }
The runtime of our exact algorithm for the DIMACS graphs is 
in most cases comparable to that of CP and higher than that of {\it cliquer}
and {\it MCQD+CS}.
For these graphs, only Pruning 2 was found to be effective (see the table in the Appendix for details), 
and thus the performance results agree with one's expectation. 
We include in the Appendix timing results on a larger collection of DIMACS graphs. 

It is to be noted that the DIMACS graphs are intended to serve as challenging test cases for the maximum clique problem, and graphs with such high edge densities and low vertex count are rather rare in practice. 
For example, most of them have between 20  to 1024 vertices with an average edge density of roughly 0.6. 
However, most real world graphs are often very large and sparse. Good examples are Internet topology graphs \cite{Faloutsos:1999:PRI:316188.316229}, the web graph \cite{kumar:extracting}, social network graphs \cite{Domingos:2001:MNV:502512.502525}, and the real-world graphs in our testbed. 

\subsubsection{The Heuristic}
\label{sec:exp-heuristic}

It can be seen that our heuristic runs several orders of magnitude faster than our exact algorithm,
while delivering either optimal or very close to optimal solution.
It gave the optimal solution on 25 out of the 30 test cases.
On the remaining 5 cases where it was suboptimal, it's accuracy ranges from 83\% to 99\% (on average 93\%).
Additionally, we run the heuristic by choosing a vertex randomly in Line \ref{maxDsel} of Algorithm \ref{alg:mClqHeu} instead of the one with the maximum degree. We observe that on average, the solution is optimal only for less than $40\%$ of the test cases compared to 83\% when selecting the maximum degree vertex.

\begin{figure}
  \centering
    \includegraphics[width=7.2cm]{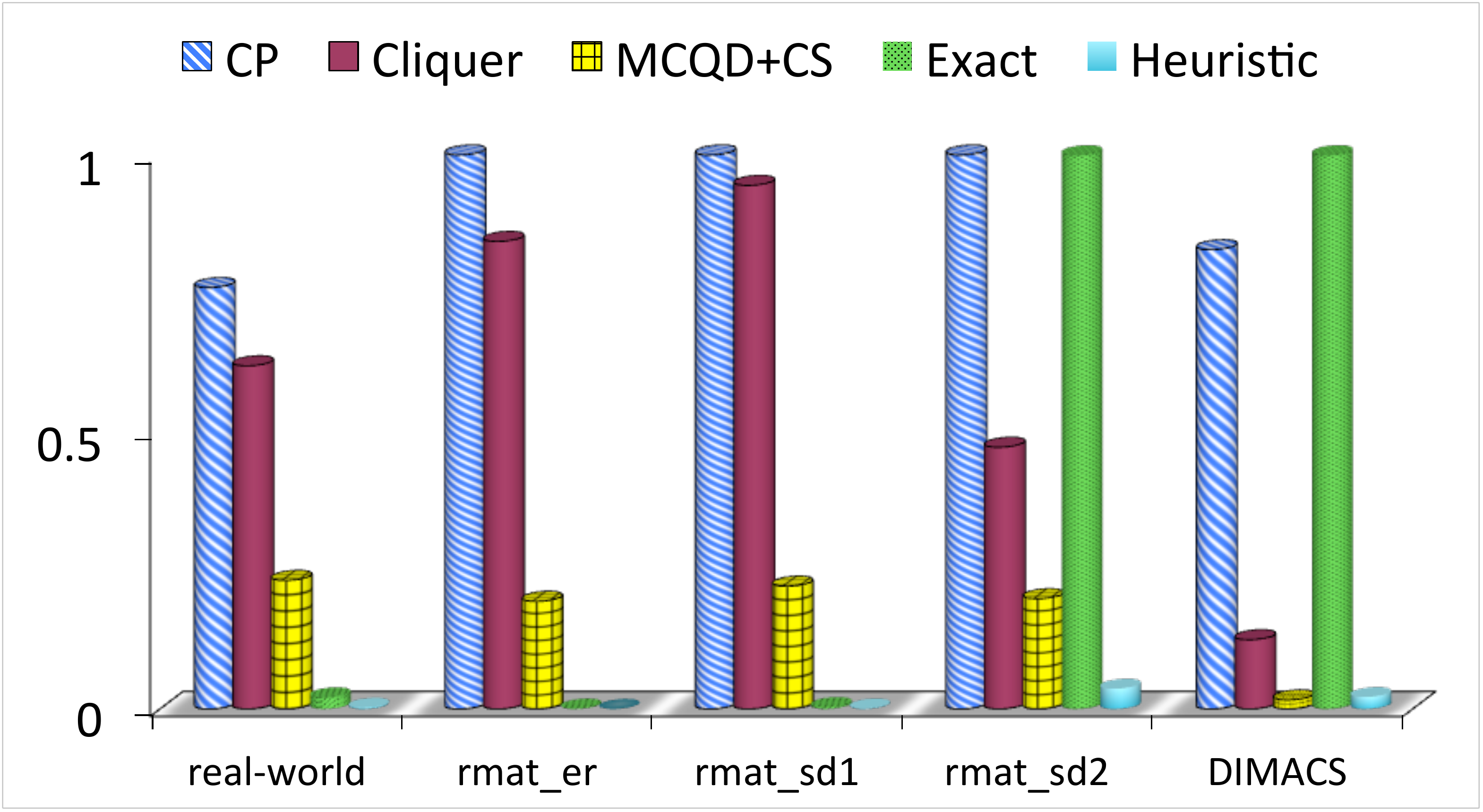}
  \caption{Runtime (normalized, mean) comparison between various algorithms. For each category of graph, first, all runtimes for each graph were normalized by the runtime of the slowest algorithm for that graph, and then the mean was calculated for each algorithm. In the various bars, graphs were considered only if the runtimes for at least three algorithms was less than the 25,000 seconds limit set.}
\label{fig-timeplot}
\end{figure}

Figure~\ref{fig-timeplot} provides an aggregated visual summary of the runtime trends of
the various algorithms across the five categories of graphs in the testbed. 

\begin{figure}
  \centering
    \includegraphics[scale=0.5]{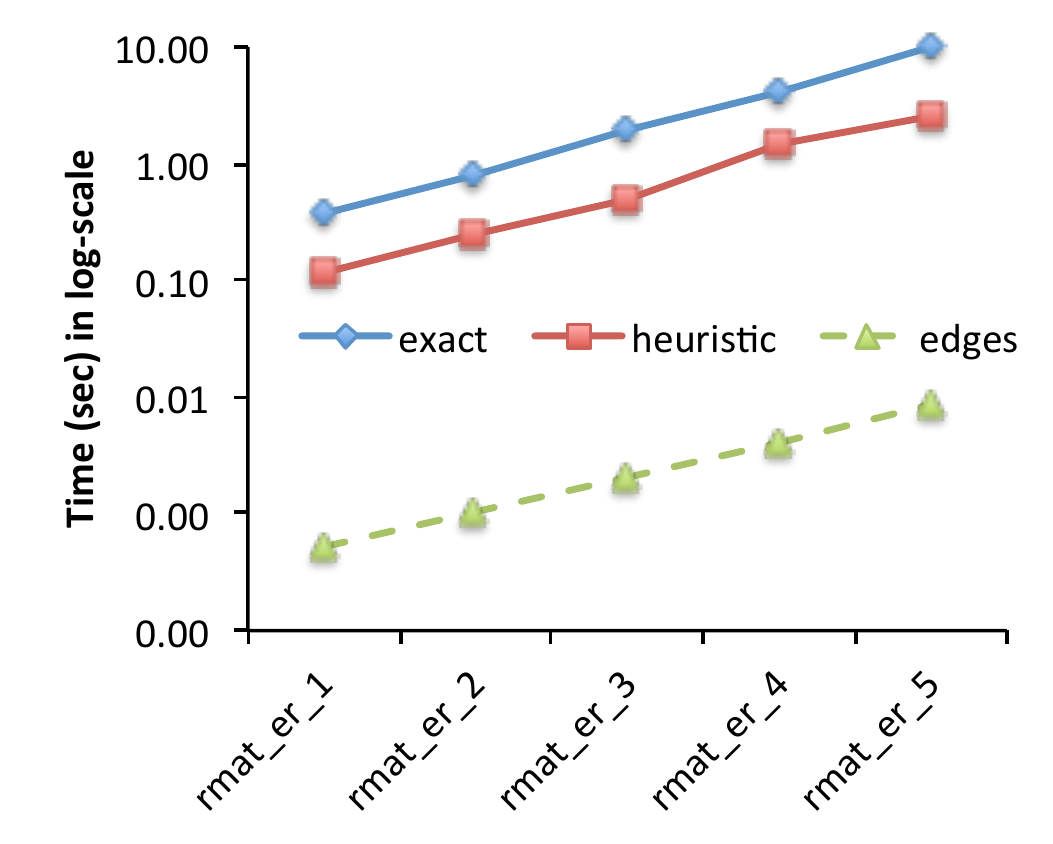}
    \includegraphics[scale=0.5]{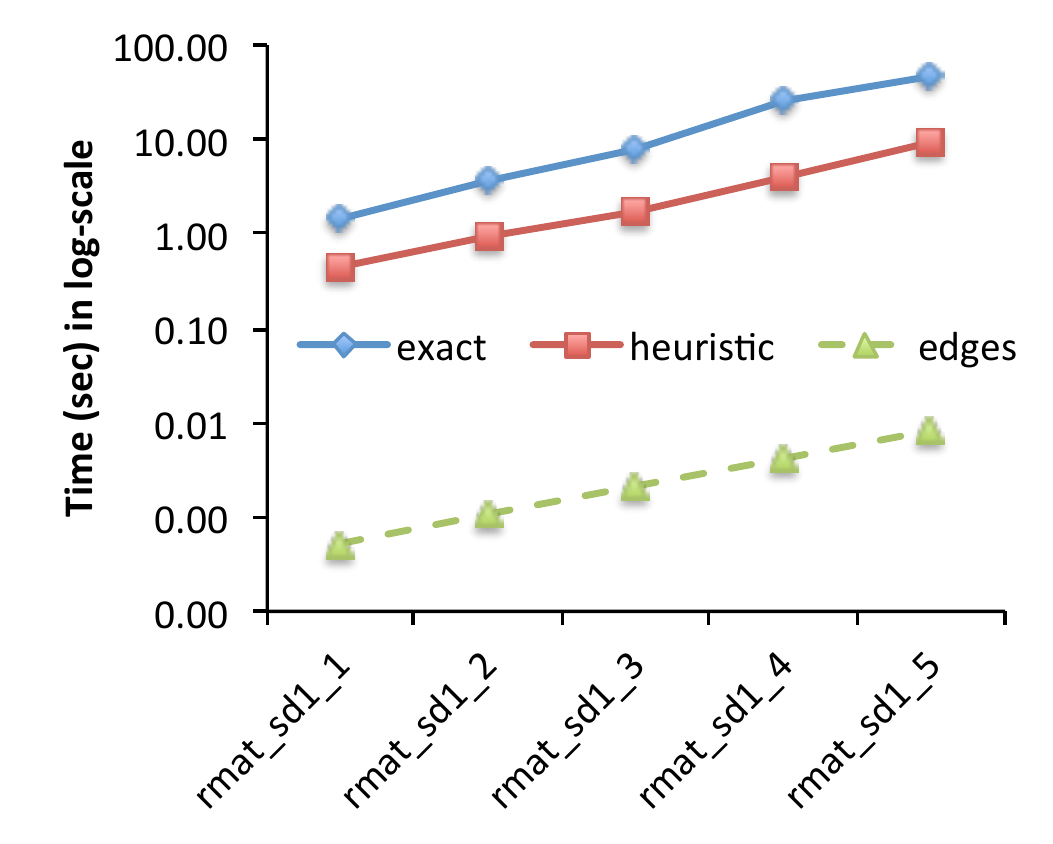}
    \includegraphics[scale=0.5]{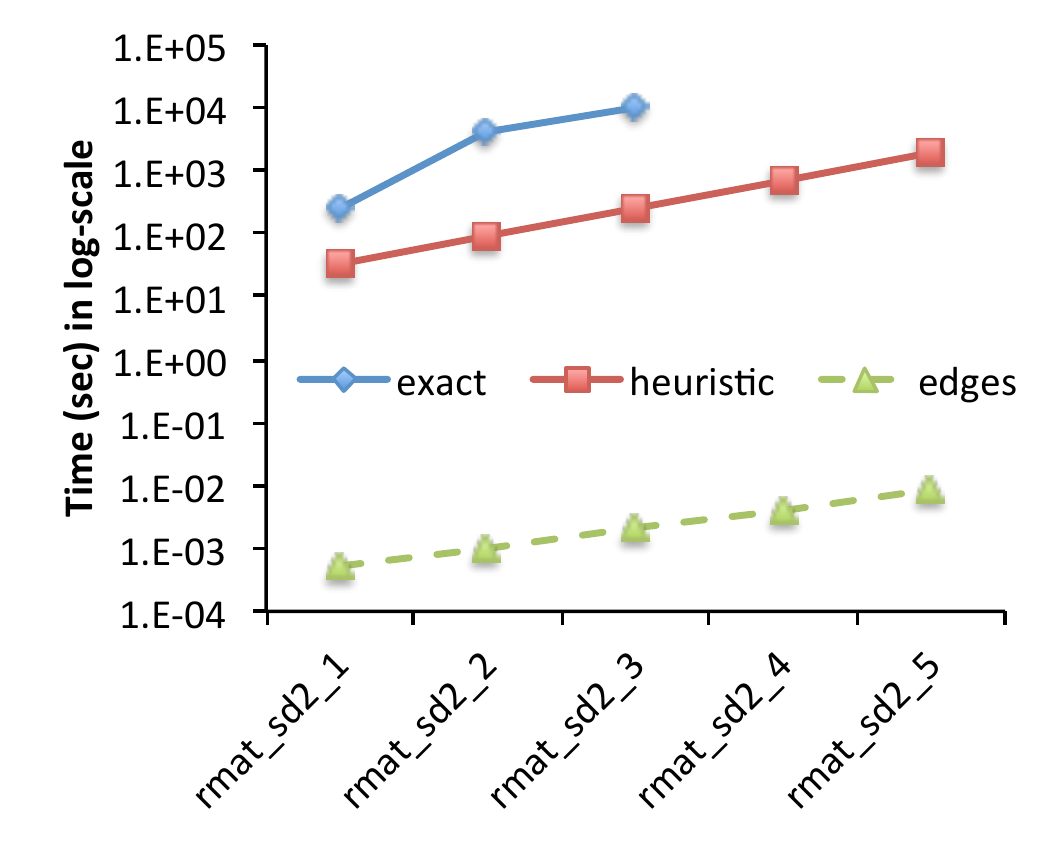}
    \includegraphics[scale=0.5]{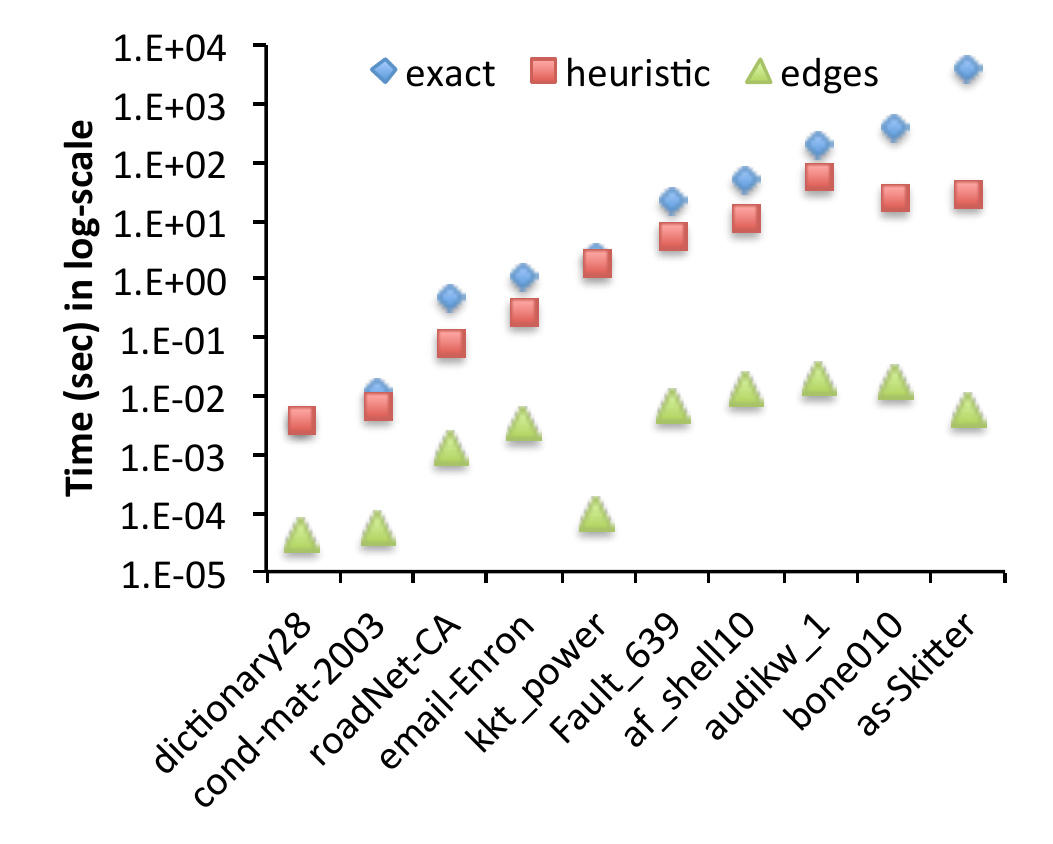}
    
 \caption{Run time plots of the new exact and heuristic algorithms. The third curve, labeled
 {\em edges}, shows the quantity number of edges in the graph divided by the clock 
 frequency of the computing platform used in the experiment. 
 }
\label{fig-runtimeplots}
\end{figure}

To give a sense of runtime growth rates, we provide in Figure~\ref{fig-runtimeplots} plots of the 
runtime of the new exact algorithm and the heuristic for the synthetic and real-world graphs 
in the testbed.  Besides the curves corresponding to the runtimes of the
{\em exact} algorithm and the {\em heuristic}, the figures also include a curve corresponding to
the number of {\em edges} in the graph divided by the clock frequency of the computing
platform used in the experiment. This curve is added to facilitate comparison between
the growth rate of the algorithms with that of a linear-time (in the size of the graph) growth rate. 
It can be seen that the runtime of the heuristic by and large grows 
somewhat linearly with the size of a graph. The exact algorithm's runtime, which is orders of
magnitude larger than the heuristic, exhibited a similar growth behavior for these test-cases
(although its worst-case complexity suggests exponential growth).

\subsection{Example of an application in social network analysis}
\label{sec:applications}

We conclude this section on experiments with a small example 
demonstrating the application of the clique algorithms for detecting overlapping communities in social networks. 
In many real networks vertices may belong to more than one group, and such groups form overlapping communities. Classical examples are social networks, where an individual usually belongs to different circles at the same time, from that of work colleagues to family, sport associations, etc. 
Finding overlapping communities is a challenging problem \cite{Fortunato_2010}.
Clique algorithms are one way in which a solution can be found.  

\begin{figure}[h!]
  \centering
    \includegraphics[width=0.9\textwidth]{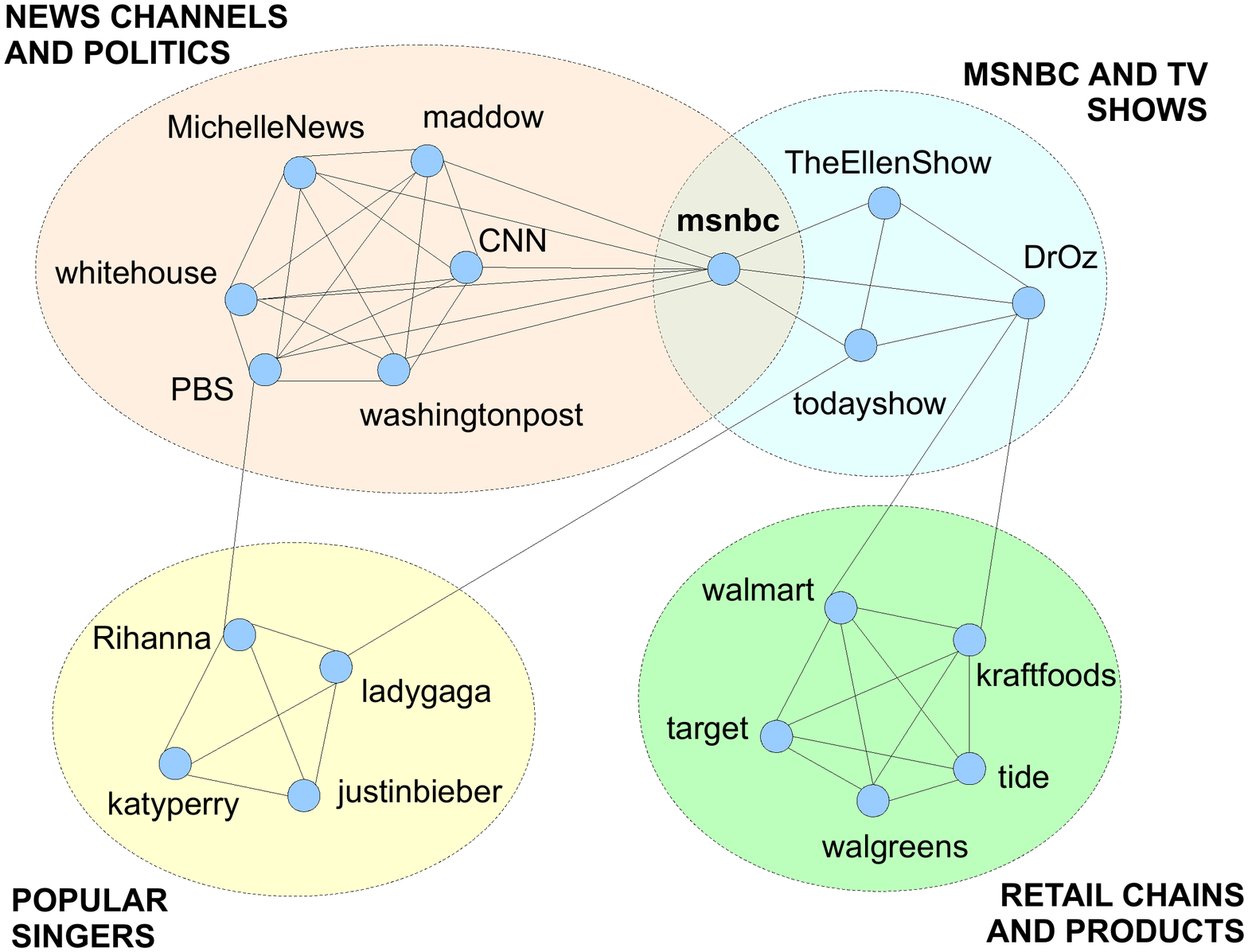}
  \caption{Some Facebook communities detected by our max clique heuristic.}
\label{fig-communities}
\end{figure}

For our small experiment, we use data collected from Facebook\footnote[1]{http://www.facebook.com}.
Every user on Facebook has a {\it wall}, which is a the user's profile space that allows the posting of messages, often short or temporal notes by other users. The user comments and user information from specific {\it walls} are publicly available and we collected them using Facebook API. We constructed a graph with the {\it walls} as vertices. Any two users who have commented on the same {\it wall} indicate a connection between the {\it walls}, and we form an edge between them. There could be many common users for each wall, and so we assigned edge weights by Jacard index or similarity coefficient \cite{Leydesdorff}. Once this is done for all {\it walls}, we retained only those edges which have weights above a chosen threshold, indicating a strong correlation. The threshold is a user's choice and decides both the size and the number of communities found.

We modified our heuristic to retain the largest maximum clique containing each node. 
The exact algorithm could have also been used instead of the heuristic for this purpose. We choose the heuristic since it is much faster and for this particular problem of community detection the accuracy of the size of cliques formed is not critical.

Figure \ref{fig-communities} shows some of the cliques/communities detected. We see two isolated communities, one for popular singers, and another for retail chains and products. We also see a community for news channels and politics, and a community of MSNBC and popular TV shows. The highlight of this experiment  is that the
clique algorithm allows a node to be a member of more than one community giving an overlapping community structure. Although the {\it news channels and politics} and {\it MSNBC and tv shows} communities are not directly related and have different members, they share a common member.


\section{Conclusion}
\label{sec:conclusion}

We presented a new exact and a new heuristic algorithm for the maximum clique problem.
We performed extensive experiments on three broad categories of graphs comparing the 
performance of our algorithms to the algorithms due to
Carraghan and Pardalos (CP) \cite{pardalos},
\"{O}sterg\.{a}rd ({\it cliquer}) \cite{ostergard} and
Konc and Janezic ({\it MCQD+CS}) \cite{konc2007improved}.
For DIMACS benchmark graphs and certain dense synthetic graphs ({\it rmat\_sd2}), our new exact algorithm performs comparably with the CP algorithm, but slower than {\it cliquer}
and {\it MCQD+CS}. 
For large sparse graphs, both synthetic and real-world, our new algorithm runs
several orders of magnitude faster than the other three. 
The heuristic, which runs many orders of magnitude faster than our exact algorithm and the others, gave optimal solution for 83\% of the test cases, and when it is sub-optimal, its accuracy ranged between 0.83 and 0.99.

The exact algorithm was in general  found to be less successful on relatively dense graphs.
An interesting line of investigation would be to study ways to overcome this.
Another line for future work would be to characterize the class(es) of graphs for which the heuristic is expected to return near-optimal solution.

\section*{Acknowledgements}
This work is supported in part by NSF award numbers CCF-0621443, OCI-0724599, CCF-0833131, CNS-0830927, IIS-0905205, OCI-0956311, CCF-0938000, CCF-1043085, CCF-1029166, and OCI-1144061, and in part by DOE grants DE-FG02-08ER25848, DE-SC0001283, DE-SC0005309, DE-SC0005340, and DE-SC0007456. 
The work of Assefaw Gebremedhin is supported by the US National Science Foundation
grant CCF-1218916.


\newpage
\appendices
\section*{Appendix}
\label{sec:appendix}

\begin{table}[!hbt]
\centering
\caption{$P1$, $P2$, $P3$, $P4$ and $P5$ are the number of vertices pruned in steps Pruning 1, 2, 3, 4, and 5 of Algorithm 1. An asterisk (*) indicates that the algorithm did not terminate within 25,000 seconds for that instance. $\omega$ denotes the maximum clique size.}
\label{tab:prunings}
\begin{tabular}{l@{\hspace{6pt}}r@{\hspace{6pt}}|@{\hspace{6pt}}r@{\hspace{6pt}}r@{\hspace{6pt}}r@{\hspace{6pt}}r@{\hspace{6pt}}r}

\toprule\toprule
$G$ 				& $\omega$ 	& 	$P1$ 		&	$P2$		& 	$P3$ 		& 	$P4$ 		&	$P5$		\vspace{-4pt} \\
				& 			&		 		& 		 		&				& 			 	&				\\ \hline \hline
{\it cond-mat-2003} 	& 	25 		& 	29,407 		&	48,096		&	6,527 		&	2,600		& 	17,576		\\
{\it email-Enron} 	& 	20 		& 	32,462 		&	155,344		&	4,060 		&	110,168		& 	8,835,739		\\ 
{\it dictionary28} 	& 	26 		& 	52,139		& 	4,353		&	2,114		&	542			& 	107			\\	
{\it Fault\_639}		&	18		&	36			&	13,987,719	&	126			&	10,767,992	&	1,116		\\
{\it audikw\_1}		&	36		&	4,101		&	38,287,830	&	59,985		&	32,987,342	&	721,938		\\
{\it bone010}		&	24		&	37,887		&	34,934,616	&	361,170		&	96,622,580	&	43,991,787	\\ 
{\it af\_shell10}		&	15		&	19			&	25,582,015	&	75			&	40,629,688 &	2,105		\\
{\it as-Skitter}		&	67		&	1,656,570		&	6,880,534		&	981,810		& 26,809,527&	737,899,486	\\ 
{\it roadNet-CA}	&	4		&	1,487,640		&	1,079,025		&	370,206		&	320,118		&	4,302		\\ 
{\it kkt\_power}		&	11		&	1,166,311		&	4,510,661		&	401,129		&	1,067,824		&	1,978,595		\\ 
\midrule
{\it rmat\_er\_1}		&	3		&	780			&	1,047,599		&	915			&	118,461		&	8,722		\\
{\it rmat\_er\_2}		&	3		&	2,019		&	2,094,751		&	2,351		&	235,037		&	23,908		\\
{\it rmat\_er\_3}		&	3		&	4,349		&	4,189,290		&	4,960		&	468,086		&	50,741		\\
{\it rmat\_er\_4}		&	3		&	9,032		&	8,378,261		&	10,271		&	933,750		&	106,200		\\
{\it rmat\_er\_5}		&	3		&	18,155		&	16,756,493	&	20,622		&	1,865,415		&	212,838		\\
\midrule
{\it rmat\_sd1\_1}	&	6		&	39,281		&	1,004,660		&	23,898		&	151,838		&	542,245		\\
{\it rmat\_sd1\_2}	&	6		&	90,010		&	2,004,059		&	56,665		&	284,577		&	1,399,314		\\ 
{\it rmat\_sd1\_3}	&	6		&	176,583		&	4,013,151		&	106,543		&	483,436		&	2,677,437		\\ 
{\it rmat\_sd1\_4}	&	6		&	369,818		&	8,023,358		&	214,981		&	889,165		&	5,566,602		\\\ 
{\it rmat\_sd1\_5}	&	6		&	777,052		&	16,025,729	&	455,473		&	1,679,109		&	12,168,698	\\ 
\midrule
{\it rmat\_sd2\_1}	&	26		&	110,951		&	853,116		&	88,424		&	1,067,824		&	614,813,037	\\ 
{\it rmat\_sd2\_2}	&	35		&	232,352		&	1,645,086		&	195,427		&			81,886,879	&	1,044,068,886	\\ 
{\it rmat\_sd2\_3}	&	39		&	470,302		&	3,257,233		&	405,856		&			45,841,352	&	1,343,563,239	\\ 
{\it rmat\_sd2\_4}	&	43		&	*			&	*			&	*			&		*		&	*			\\
{\it rmat\_sd2\_5}	&	N		&	*			&	*			&	*			&		*		&	*			\\
\midrule
{\it hamming6-4}	&	4		&	0			&	704			&	0			&	583			&	0			\\
{\it johnson8-4-4}	&	14		&	0			&	1855			&	0			&	136,007		&	0			\\
{\it keller4}		&	11		&	0			&	9435			&	0			&	8,834,190		&	0			\\
{\it c-fat200-5}		&	58		&	0			&	8473			&	0			&	70449		&	0			\\
{\it brock200\_2}	&	12		&	0			&	9876			&	0			&	349,427		&	0			\\
\bottomrule\bottomrule
\end{tabular}
\end{table}

\begin{table}[tbh]
\small
\centering
\caption{\footnotesize Comparison of runtimes of algorithms \cite{pardalos} ({\it CP}), 
\cite{ostergard} ({\it cliquer}) and \cite{konc2007improved} ({\it MCQD+CS})
with that of our new exact algorithm ($\tau_{new-exact}$) for DIMACS graphs. 
An asterisk (*) indicates that the algorithm did not terminate within 
10,000 seconds for that instance. $\omega$ denotes the maximum clique size, $\omega_{new-heuristic}$ the maximum clique size found by our heuristic and $\tau_{new-heuristic}$, its runtime.}

\label{tab:dimacs}
\begin{tabular}{l@{\hspace{6pt}}r@{\hspace{6pt}}r@{\hspace{6pt}}r@{\hspace{6pt}}|@{\hspace{6pt}}r@{\hspace{6pt}}r@{\hspace{6pt}}r@{\hspace{6pt}}r@{\hspace{6pt}}|@{\hspace{6pt}}r@{\hspace{3pt}}r}
\toprule\toprule
& &     &   &  & &$\tau_{MCQD}$&$\tau_{new-}$ & $\omega_{new-}$ & $\tau_{new-}$\\
$G$        &   $\left|V\right|$    &   $\left|E\right|$    &   $\omega$    &$\tau_{CP}$&$\tau_{cliquer}$&$_{+CS}$&$_{exact}$ & $_{heuristic}$ & ${heuristic}$\\ \hline \hline
{\it brock200\_1}	&	200		&	14,834	&	21	&	*		&	10.37	&	0.75		&	*		&	18	&	0.02	\\
{\it brock200\_2}	&	200		&	9,876	&	12	&	0.98		&	0.02		&	0.01		&	1.1		&	10	&	$<$0.01	\\
{\it brock200\_3}	&	200		&	12,048	&	15	&	14.09	&	0.16		&	0.03		&	14.86	&	12	&	$<$0.01	\\
{\it brock200\_4}	&	200		&	13,089	&	17	&	60.25	&	0.7		&	0.12		&	65.78	&	14	&	$<$0.01	\\
{\it c-fat200-1}		&	200		&	1,534	&	12	&	$<$0.01	&	$<$0.01	&	$<$0.01	&	$<$0.01	&	12	&	$<$0.01	\\
{\it c-fat200-2}		&	200		&	3,235	&	24	&	$<$0.01	&	$<$0.01	&	$<$0.01	&	$<$0.01	&	24	&	$<$0.01	\\
{\it c-fat200-5}		&	200		&	8,473	&	58	&	0.6		&	0.33		&	0.01		&	0.93		&	58	&	0.04	\\
{\it c-fat500-1}		&	500		&	4,459	&	14	&	$<$0.01	&	$<$0.01	&	$<$0.01	&	$<$0.01	&	14	&	$<$0.01	\\
{\it c-fat500-2}		&	500		&	9,139	&	26	&	0.02		&	$<$0.01	&	0.01		&	0.01		&	26	&	0.01	\\
{\it c-fat500-5}		&	500		&	23,191	&	64	&	3.07		&	$<$0.01	&	$<$0.01	&	*		&	64	&	0.11	\\
{\it hamming6-2}	&	64		&	1,824	&	32	&	0.68		&	$<$0.01	&	$<$0.01	&	0.33		&	32	&	$<$0.01	\\
{\it hamming6-4}	&	64		&	704		&	4	&	$<$0.01	&	$<$0.01	&	$<$0.01	&	$<$0.01	&	4	&	$<$0.01	\\
{\it hamming8-2}	&	256		&	31,616	&	128	&	*		&	0.01		&	0.01		&	*		&	128	&	0.67	\\
{\it hamming8-4}	&	256		&	20,864	&	16	&	*		&	$<$0.01	&	0.1		&	*		&	16	&	0.03	\\
{\it hamming10-2}	&	1,024	&	518,656	&	512	&	*		&	0.31		&	-		&	*		&	512	&	95.24	\\
{\it johnson8-2-4}	&	28		&	210		&	4	&	$<$0.01	&	$<$0.01	&	$<$0.01	&	$<$0.01	&	4	&	$<$0.01	\\
{\it johnson8-4-4}	&	70		&	1,855	&	14	&	0.19		&	$<$0.01	&	$<$0.01	&	0.23		&	14	&	$<$0.01	\\
{\it johnson16-2-4}	&	120		&	5,460	&	8	&	20.95	&	0.04		&	0.42		&	22.07	&	8	&	$<$0.01	\\
{\it keller4}		&	171		&	9,435	&	11	&	22.19	&	0.15		&	0.02		&	23.35	&	11	&	$<$0.01	\\
{\it MANN\_a9}		&	45		&	918		&	16	&	1.73		&	$<$0.01	&	$<$0.01	&	2.5		&	16	&	$<$0.01	\\
{\it MANN\_a27}	&	378		&	70,551	&	126	&	*		&	*		&	3.3		&	*		&	125	&	1.74	\\
{\it p\_hat300-1}	&	300		&	10,933	&	8	&	0.14		&	0.01		&	$<$0.01	&	0.14		&	8	&	$<$0.01	\\
{\it p\_hat300-2}	&	300		&	21,928	&	25	&	831.52	&	0.32		&	0.03		&	854.59	&	24	&	0.03	\\
{\it p\_hat500-1}	&	500		&	31,569	&	9	&	2.38		&	0.07		&	0.04		&	2.44		&	9	&	0.02	\\
{\it p\_hat500-2}	&	500		&	62,946	&	36	&	*		&	159.96	&	1.2		&	*		&	34	&	0.14	\\
{\it p\_hat700-1}	&	700		&	60,999	&	11	&	12.7		&	0.12		&	0.13		&	12.73	&	9	&	0.04	\\
{\it p\_hat1000-1}	&	1,000	&	122,253	&	10	&	97.39	&	1.33		&	0.41		&	98.48	&	10	&	0.11	\\
{\it san200\_0.7\_1}	&	200		&	13,930	&	30	&	*		&	0.99		&	$<$0.01	&	*		&	16	&	0.01	\\
\bottomrule
\bottomrule
\end{tabular}
\end{table}

\end{document}